\documentstyle[
	aps,
	twocolumn
	]{revtex}

\begin{document}


\title{Models for adatom diffusion on FCC(001)
metal surfaces}
\author{Hanoch Mehl, Ofer Biham and Itay Furman}
\address{
Racah Institute of Physics,
The Hebrew University,
Jerusalem 91904,
Israel}
\author{Majid Karimi}
\address{
Physics Department,
Indiana University of Pennsylvania,
Indiana, PA 15705
}
\maketitle

\begin{abstract}
We present a class of
models that describe self-diffusion
on FCC(001) metal substrates
within a common framework.
The models are tested for
Cu(001), Ag(001), Au(001), Ni(001) and Pd(001), 
and 
found to apply well for all of them.
For each of these metals 
the models can be used to estimate the activation
energy of any diffusion process using a few basic parameters
which may be obtained from experiments, ab-initio 
or semi-empirical calculations.
To demonstrate the approach,
the parameters of the models are optimized to describe
self-diffusion on the (001) surface, 
by comparing the energy barriers
to a full set of barriers obtained 
from semi-empirical potentials
via the embedded atom method (EAM).  
It is found that these models 
with at most four parameters,
provide a good description of the full landscape
of hopping energy barriers
on FCC(001) surfaces. 
The main features of the diffusion processes 
revealed by EAM calculations are quantitatively
reproducible by the models. 
\end{abstract}


\section{Introduction}

Thin film growth processes involve 
complicated kinetics 
giving rise to a rich variety of surface morphologies.
Within this vast domain, the study of the growth in the 
submonolayer regime is of particular interest due to the
large impact of the initial kinetics on the resulting
film structure.
Experiments on thin film growth on well characterized substrates
using molecular-beam epitaxy (MBE) have provided a large body of information
about growth kinetics and morphology,
and revealed that for a variety of systems and a broad temperature range,
island nucleation is the dominant mechanism for crystal growth
\cite{Kunkel1990,Esch1994}.
Diffraction methods such as helium beam scattering 
\cite{Kunkel1990,Ernst1992,Ernst1992a,Li1993,Ernst1994},
low energy electron diffraction
\cite{Zuo1991,zuo1994,Durr1995,Swan1997}
and other techniques
\cite{Breeman1992,Stroscio1993},
provide information on the collective behavior
and the statistical properties of the surface.
These techniques have been used to measure
the island size distribution, the island density, and
their scaling properties with respect to the coverage and the flux
\cite{Ernst1992,Zuo1991,zuo1994,Durr1995,Swan1997}.
The variation of the island density
with respect to the temperature was also studied
\cite{Ernst1992,Durr1995,Breeman1992}.

More detailed information at the atomic scale is provided by
scanning tunneling microscopy (STM)
\cite{Esch1994,Stroscio1993,Hwang1991,Mo1991,Potschke1991,Bott1992,Kopatzki1993,Michely1993,Roder1993,Girard1994,Gunther1994,Stroscio1994,Vrijmoeth1994,Bromann1995,Roder1995,Linderoth1996,Wen1994,Meyer1995,Morgenstern1995,Morgenstern1996,Pai1997}.
Most notably, STM provides means to study the variety of
morphologies encountered in the different systems,
or in the same system under different growth conditions
\cite{Esch1994,Stroscio1993,Hwang1991,Mo1991,Potschke1991,Bott1992,Kopatzki1993,Michely1993,Roder1993,Girard1994,Vrijmoeth1994}.
In some experiments STM was used to 
acquire information on larger scales, e.g., island size distributions
\cite{Stroscio1993,Mo1991,Girard1994,Gunther1994,Stroscio1994,Linderoth1996}.
Despite the wealth of experimental results at the atomic scale,
for decay rates of small islands, mobility of small islands and
edge diffusion 
\cite{Gunther1994,Vrijmoeth1994,Bromann1995,Roder1995,Linderoth1996,Wen1994,Morgenstern1995,Pai1997},
the underlying energetics is mostly 
inaccessible to direct experimental measurements.
Thus, one must rely on theory to extract
activation energies from the experimental results,
and these are usually limited in number,
and sometimes are subject to alternative interpretations.

The only technique which provides direct access
to diffusion processes and activation energies at the atomic scale is
field ion microscopy (FIM)
\cite{Chen1990,Kellogg1990,Kellogg1991,Wang1993,Kellogg1994}.
This technique was used to identify the diffusion modes of adatoms
\cite{Chen1990,Kellogg1990}
as well as small islands 
\cite{Kellogg1991,Kellogg1994}
on FCC(001) metal surfaces,
to measure their diffusion coefficients,
and to determine the sticking process of adatoms to an island
\cite{Wang1993}.
Recently there were several attempts to use STM
to derive such local information directly
\cite{Bromann1995}.

Theoretical studies aimed at providing better understanding of
the relation between key processes at the atomic scale and the
resulting morphologies have been done using Monte Carlo (MC)
simulations
\cite{Bartelt1992,Bartelt1993,Zhang1993,Amar1994,Bales1994,Barkema1994,Breeman1994,Jensen1994,Ratsch1994,Wolf1994,Zhang1994,Schroeder1995,Amar1995,Ratsch1995,Bales1995,Amar1996,Breeman1996,Furman1997,Voter1986,Clarke1988,Jacobsen1995}.
In simulations of island growth during deposition, 
atoms are deposited randomly on the substrate 
at rate $F$ [given in monolayers (ML) per second]
and then hop, attach to, and detach from existing islands
according to some model.
A common approach is to assume some key processes
and their rates, and then simulate the growth process
\cite{Bartelt1992,Bartelt1993,Amar1994,Jensen1994}.
In some cases information such as diffusion length,
typical distance and time between nucleation events
is assumed to be known a-priori, and is put by hand into the simulation
in order to accelerate the computation
\cite{Wolf1994,Schroeder1995,Biehl1998}.
The advantage of this approach is that the models are well
defined 
and use only few parameters.
These models are useful for studies of scaling and
morphology but cannot provide a quantitative description
of diffusion on a particular substrate.
Furthermore, they account only
for a limited number of processes, that are assumed to
be the only significant ones.

A complementary scheme employs the underlying activation energies.
In this scheme the hopping rate $h$ (in units of hops per second)
of a given atom to each unoccupied nearest neighbor (NN)
site is given by 
\begin{equation}
h=\nu \cdot \exp(-E_B/k_BT)
\label{hoppingrate}
\end{equation}
where 
$\nu = 10^{12}$ $s^{-1}$ 
is the commonly used attempt rate,
$E_B$ is the activation energy barrier, 
$k_B$ is the Boltzmann constant and $T$ is the temperature.

The activation energy barrier $E_B$ depends on the local environment of the
hopping atom, namely the configuration of occupied and
unoccupied adjacent sites.
Two approaches have been taken in the construction of
the energy barriers for hopping in the simulations.
One approach was to construct simple models that include
the desired features, such as stability and mobility of 
small islands, and  that take into account properties such as bond
energies
\cite{Bales1994,Ratsch1994,Zhang1994,Amar1995,Ratsch1995,Bales1995,Amar1996,Furman1997,Clarke1988}.
In general, this approach encompasses both the virtues and the drawbacks
of the simpler approach presented before.

A second approach is based on the use of an approximate many-body 
energy functional to calculate the hopping energy
barriers for a complete set of relevant configurations
\cite{Zhang1993,Barkema1994,Breeman1994,Breeman1996,Voter1986,Jacobsen1995}.
This approach provides a good description of diffusion
processes on the given substrate but only limited 
understanding due to the large number of parameters.

In this paper we extend and further explore a framework for
a {\it systematic} derivation of simple models for self-diffusion
on FCC(001) surfaces out of a detailed and complicated
set of energy barriers.
Simple in this context means
that only a small number of parameters are involved
and all have a definite and intuitive interpretation.
Using sensible assumptions about the bond energies and 
diffusion paths we obtain simple formulae for the activation energy
barriers. 
We then {\it optimize} the parameters of these formulae 
for each metal separately
by using energy barriers 
obtained from the embedded-atom method.
This procedure gives rise to simple models
that have at most four parameters and provide
good quantitative description of the landscape of
hopping energy barriers.
In a previous publication we have introduced the framework
and applied it to Cu/Cu(001) growth
\cite{Biham1998}.
Here we make a three-fold step forward.
(a) We include four other FCC metals in the model
and derive the appropriate parameters for them.
This shows the utility of the models
and provide a unifying framework 
that applies to a large class of metals;
(b) Non-linear interactions are introduced in addition
to the linear interactions considered before.
This allows for a more accurate optimization without increasing
the number of parameters;
(c) We explore the basis of the assumptions underlying this scheme
and the extent of their applicability.

The paper is organized as follows. 
In Sec. II we introduce the physical framework of the model
and discuss its underlying assumptions.
The results of the EAM calculations are given in Sec. III.
The models are introduced in Sec. IV 
with the fitting to the EAM results.
This is followed by a discussion of
the results and their implications in Sec. V.

\section{Applicability considerations}
\label{app_section}

The framework developed in this paper assumes several
characteristics of the diffusion processes considered.
It applies to systems for which these
assumptions are valid,
which includes most of the FCC metals in
the moderate temperature regime $(\approx200-500 K)$.
The following assumptions are employed throughout the discussion.

\noindent
{\em  Bridge-site hopping of adatoms is in general dominant
over exchange hopping} (Fig. \ref{fig:terrace_diffusion}).
There has been a controversy concerning this assumption. 
Using semi-empirical methods
the barrier for exchange hopping for Cu(001) was estimated 
to be $0.2$ eV
\cite{Hansen1991,Hansen1993,Perkins1993},
in agreement with the experimental data that was available
at that time
\cite{Ernst1992}.
This is much lower than the barrier for bridge-site hopping. 
However, theoretical work using several other methods  
\cite{Liu1991,Boisvert1997,Karimi1995} 
including EAM
indicate that the exchange barrier for Cu is much higher (more than $0.8$ eV)
and therefore bridge-site hopping is dominant.
This conclusion
is also supported by recent experimental work and reinterpretation of the
previous findings
\cite{Durr1995}.
In general, bridge-site hopping is found to be dominant in
all the metals studied here.
(for Au, some exchange processes appear to be significant,
yet in most cases bridge hopping is favorable). 
Due to the exponential dependence of
the rate of each process on the corresponding energy barrier, 
it is generally reasonable to 
take into account only the mechanism which is energetically favorable
(bridge-site hopping, in this case)
and neglect the mechanism which exhibits higher activation energy barrier
(exchange mechanism).

\noindent
{\em Only nearest and next-nearest neighbor interactions are significant}.
Within this assumption one can obtain the activation energies
for most diffusion processes to a good accuracy. 
However there are some processes such as vacancy
diffusion, where a larger
environment affects the diffusing atom. 

\noindent
{\em  There is one common attempt frequency for all processes.} 
Since there is no systematic knowledge about the 
dependence of the attempt frequency on the local environment,
the assumption of one common frequency is the usual practice.
Estimations for
the attempt frequency can be obtained using MD simulations 
\cite{Boisvert1997}.
Another way is to fit molecular static (MS) data to harmonic potential to
find an effective
force constant, and then deduce the frequency of oscillations. 
So far,
little work has been 
done on this subject. Previous works, including interpretation of
experimental results, usually pre-suppose some common attempt frequency 
in the range $10^{12}$--$10^{13}s^{-1}$\cite{Allnatt1993}.

\section{The EAM barriers}
\label{eam_section}

The models described in this work are tested
by fitting their parameters to
energy barriers for self-diffusion
of Cu(001), Ag(001), Au(001), Ni(001) and Pd(001) surfaces, 
obtained using EAM
\cite{Daw1983}.
This method uses 
semi-empirical potentials and  provides a good 
description of
self-diffusion on such surfaces
\cite{Karimi1995}.
Specifically, for
all the metals considered here the EAM functions developed by Adams,
Foiles, and Wolfer (AFW) 
\cite{Adams1989}
are employed. 
These functions are fitted
to a similar but more accurate data base as the one employed by 
Foiles, Baskes, and Daw
\cite{Foiles1986}.
The calculations are done on a slab of 20 square 
layers with 100 atoms in each layer.

When an atom on the surface hops into a vacant nearest neighbor site
it has to cross the energy barrier between the initial and final
sites. 
We have used molecular statics in conjunction with the EAM functions
to find that energy barrier. 
This is simply the difference
between the energy at the bridge site 
(or more precisely, at the point along the path with highest energy)
and in the initial site. 

The hopping energy barriers are calculated for all local environments
as shown in Fig. \ref{fig:env}, where 
seven adjacent sites, 
$i=0,\dots,6$ are taken into account, according to the 
assumptions presented in Sec. II.
Each one of these sites can be either 
occupied ($S_i=1$) or vacant ($S_i=0$), 
giving rise to $2^7=128$ barriers. 
A binary representation is used
to assign indices to these barriers. 
For each configuration $(S_0,\dots,S_6)$ the barrier is 
given by $E_B^n$,
where

\begin{equation}
n=\sum_{i=0}^6 S_i \cdot 2^i
\label{barindex}
\end{equation}

\noindent
takes the values
$n=0,\dots,127$. 
The full set of hopping energy barriers (given in eV)
is presented in Table \ref{eam_table},
for  Cu(001), Ag(001), Au(001), Ni(001) and Pd(001).
To show these values in a compact form, 
each barrier in Table I corresponds to a
configuration in which the occupied sites are the union of the
occupied sites in the picture on top of the given column and
on the left hand side of the given row. 
The column in Table I in which a given configuration appears
is determined by the occupancy of sites
$i=2,3,6$
while the row is determined by sites
$i=0,1,4,5$.
One can define

\begin{equation}
n_1 = \sum_{i=2,3,6} S_i \cdot 2^i; \ \ \ n_2 = \sum_{i=0,1,4,5} S_i \cdot 2^i
\label{n1n2}
\end{equation}

\noindent
such that for each configuration 
$n=n_1 + n_2$.
To demonstrate the use of Table \ref{eam_table} ,
 we will check for Cu(001) the barrier of the 
configuration in which sites $0,3$ and $4$ are occupied and all
other sites adjacent to the hopping atom are vacant.
For this configuration, according to Eq. 
(\ref{n1n2}),
$n_1=8$ and $n_2=17$ ($n=25$).
The barrier, that is found in the column with the
index $8$ and the row with index $17$,
in the line of Cu,
is 
$E_B^{25} = 0.89$  eV.

In Table  \ref{eam_table}
we use the symmetries of the configurations in the
$3 \times 3$ cell (Fig. \ref{fig:env})
to reduce the number of entries. 
There is a mirror symmetry plane perpendicular to the surface and 
containing the arrow of the 
hopping atom. 
Consequently, the columns of $n_1=4$ and $12$,
in which site $i=2$ is occupied, stand also for the 
symmetric configurations in which 
$i=6$ is occupied.
In the other four columns,
there are some configurations that, due to symmetry,
appear twice. 
In such cases, the barrier for the configuration with larger
$n$ appears in {\it italics}.

For the purpose of the calculations and parameterization of
the model
we consider only hopping moves in which a single atom  
hops each time.
It turns out, however, that in some cases the molecular statics calculations,
used to obtain the barriers, give rise to concerted moves. 
In such moves the atom at site $i=3$ follows the hopping atom
and takes the  place vacated by the hopping atom. 
This fact significantly reduces the barrier.
It turns out that for configurations in which concerted
moves appear, they can be suppressed by adding a column
of three atoms on the left hand side of sites
$i=0,3$ and $4$.
In Table I, the energy values for those  configuration in which a concerted 
move was found, 
are shown in parenthesis. 
The barrier
obtained when the concerted move was suppressed 
is shown to the left of the parenthesis.

To gain a better understanding of the barrier energy 
landscape we present the
barrier height distribution 
(without concerted moves)
in Fig. \ref{histogram} for the five metals considered.
We observe that this distribution exhibits four groups.
This feature is in agreement with Ref. 
\cite{Breeman1994}
where a different method 
\cite{Finnis1984}
was used to calculate the barriers. 
Each group, corresponds to a single or a double column
in Table I.
In general, group I includes very fast moves towards island edges,
group II includes moves along the edge, group III includes, most notably,
the single atom move ,while group IV includes detachment 
moves.

\section{The Models}

\subsection{The Additivity Assumption}
\label{linearity}

The starting point in the construction of
a simple model that describes the
hopping energy barriers for all the configurations of 
Fig. \ref{fig:env},
is the assumption that the
contributions of all adjacent atoms to the 
energy barrier add up linearly.
To examine this assumption for Cu(001), we 
evaluated directly the binding energies 
within the EAM approach
for a series of configurations from which we extracted the
relevant bond energies. 

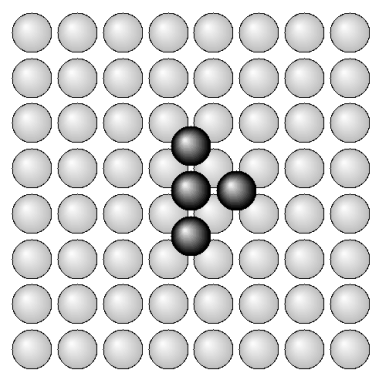
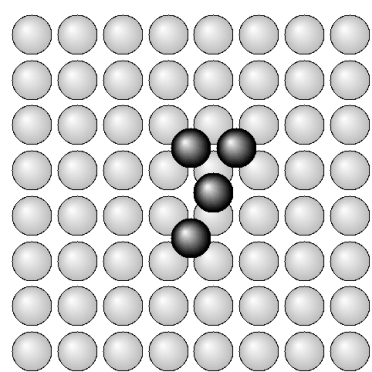
\begin{figure}
\caption{Configurations used to evaluate the NN bond energy:
(a) two nearest neighbors of an adatom may be next nearest
neighbors of each other; (b) two nearest neighbors of an atom at the bridge
site, may also be nearest neighbors of each other.}
\label{linearity2}
\end{figure}

The binding energy of a given configuration of adatoms on the surface
is evaluated as follows: 
First we calculate the total energy of the system in that configuration.
Then we find the total energy of another configuration
in which there is the same number of adatoms on the surface
but they are far apart from each other.
(by that we mean that moving them one lattice site in any direction
would not change the total energy). 
The binding energy between the adatoms is given
by the difference in
the total energies between the two configurations. 
An example of the 
procedure is shown in
Fig. \ref{linearity1}.
It appears, that 
the evaluation of the NNN bond energy is easier since 
one can construct a sufficiently large set of 
configurations which include only
NNN bonds
with no NN bonds.
In the case of NN bonds,
most of the relevant configurations 
also include NNN bonds [Fig. \ref{linearity2} (a)]. 
Similarly, considering an atom on top of a bridge site,
typical configurations which include atoms adjacent to
the bridge site exhibit NN bonds between them
as shown in 
Fig. 
\ref{linearity2}(b).

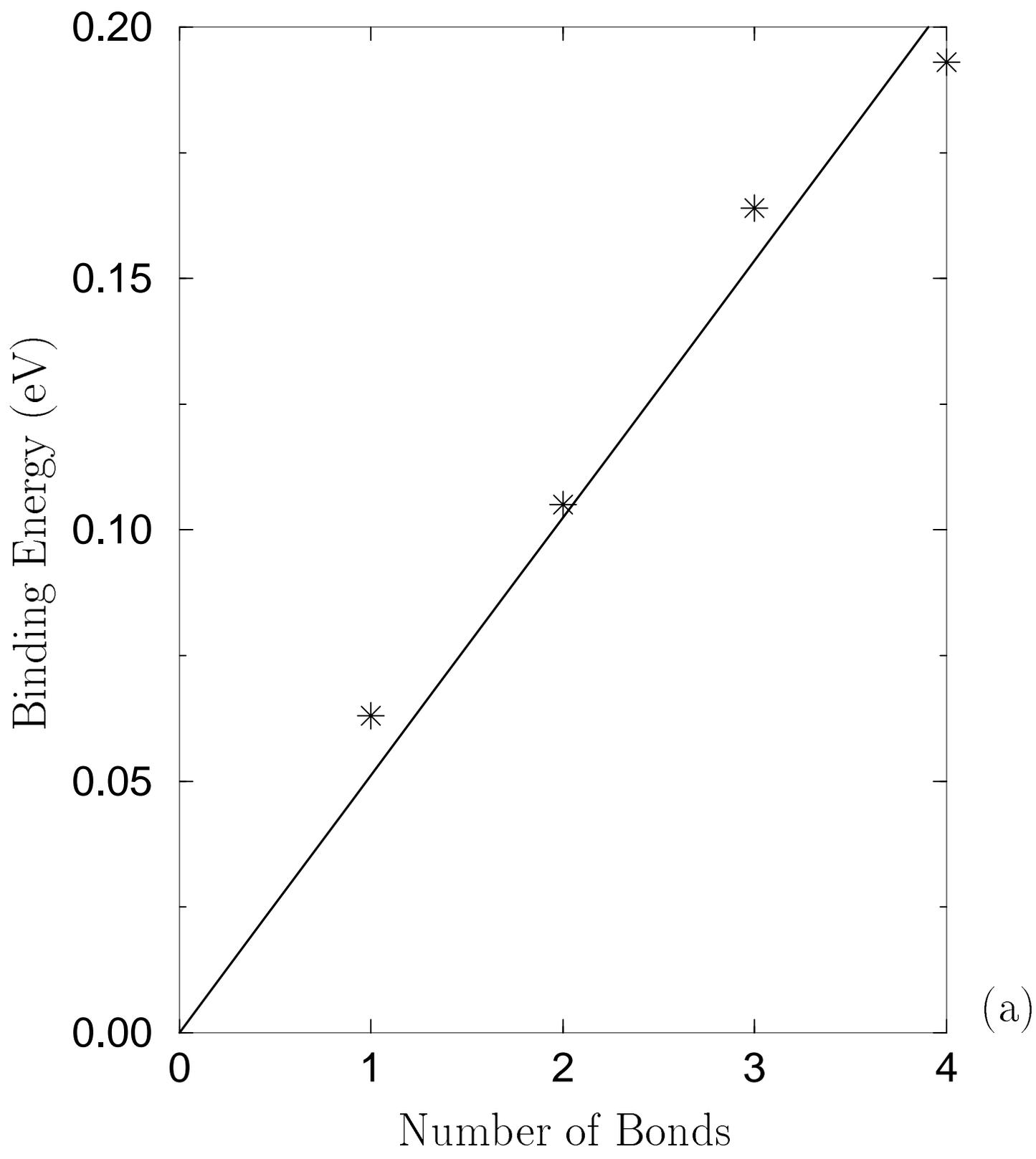
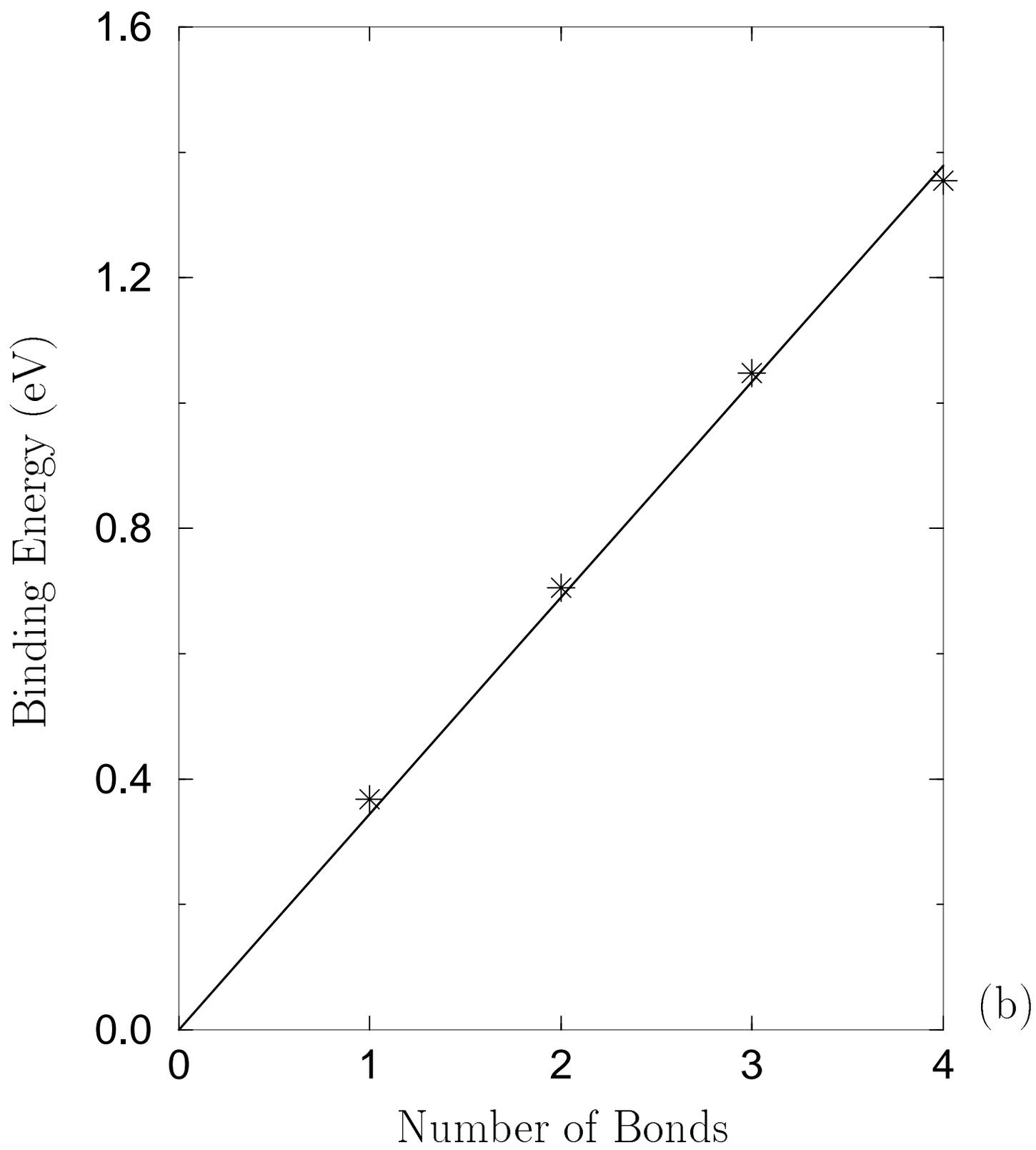
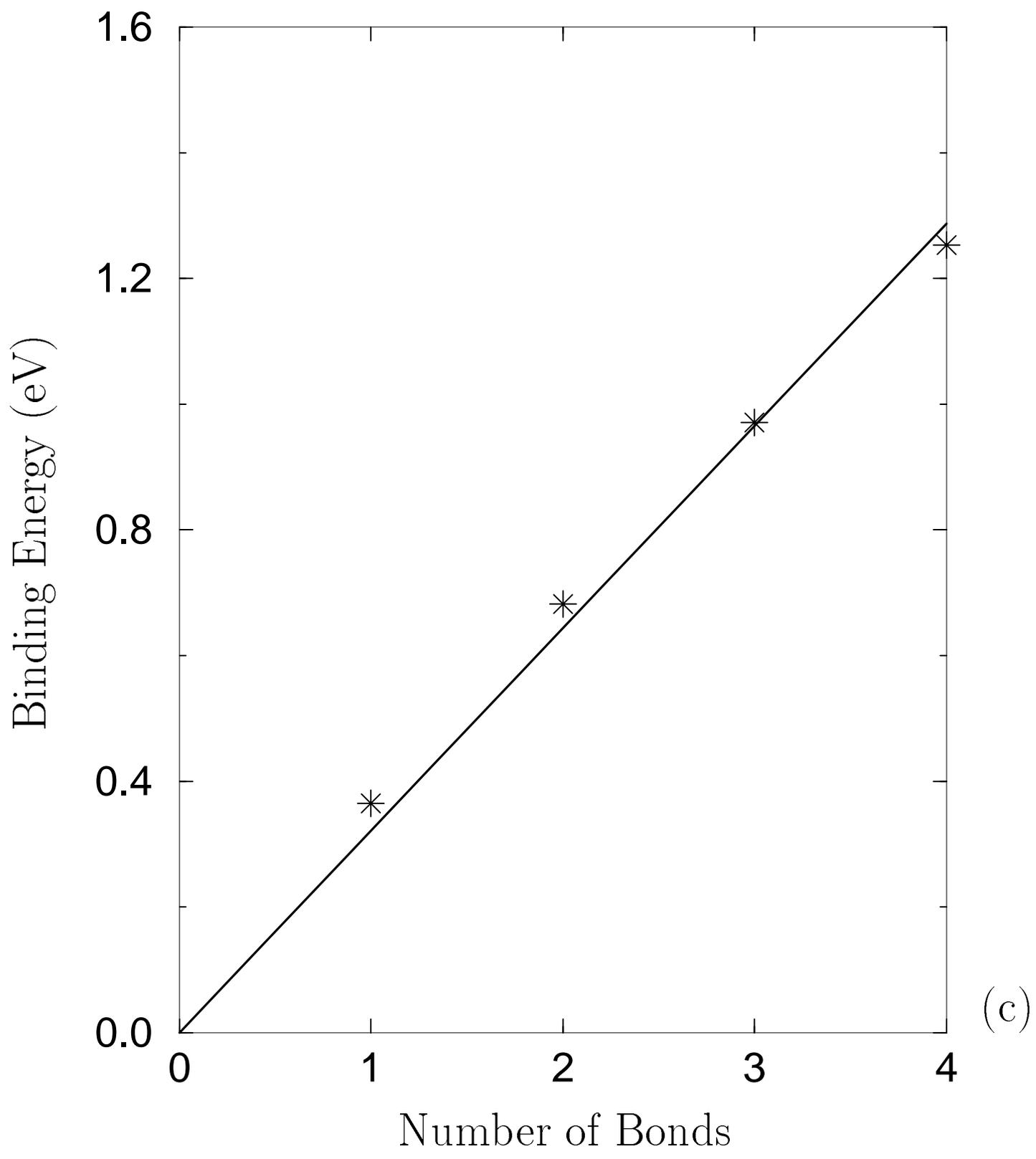
\begin{figure}
\caption{Testing the additivity assumption for 
bond energies on the Cu(001) surface.
(a) binding energy vs. the number of NNN bonds. The best
linear fit yields a value of $E_{NNN}=0.0512 eV$; 
(b) binding energy vs. number of NN bonds, where 
the binding Energy of NNN bonds is subtracted.
The slope of the solid line yields $E_{NN}= 0.324 eV$; 
(c) binding energy vs. number of NN bonds for an atom 
at the bridge site, where other NN
bond energies are subtracted. 
The best fit is $E_{top}= 0.345 eV$.
The fits include a constraint that the line passes through the origin,
since an isolated adatom has zero adatom-adatom binding energy.}
\label{linear:graph}
\end{figure}

Therefore, we will first examine the additivity of the NNN bonds
employing a series of four configurations in which an adatom
has 1, 2, 3 and 4 NNN on the surface. For each one of these 
four configurations the total energy is compared to that of
a configuration with the same number of adatoms,
in which they are far apart from each other.
In Fig.
\ref{linear:graph}(a),
the total binding energy between adatoms is plotted as a function
of the number of NNN bonds. The best linear fit is drawn, and its
slope yields a value of
$E_{NNN} = 0.0512$ eV.
The next step is to examine the linearity of the NN binding energy
using a series of four configurations in which an adatom
has 1, 2, 3 and 4 NN's on the surface, within a similar procedure. 
The NNN bonds in each configuration are deducted using the value
obtained before. The results are shown in 
Fig. 
\ref{linear:graph}(b)
and the NN bond energy is obtained:
$E_{NN} = 0.324$ eV. 
A similar analysis for an adatom on a bridge site is shown 
in Fig. 
\ref{linear:graph}(c)
and the binding energy between an atom on the bridge site and
an adjacent atom is given by
$E_{NN(bridge)} = 0.345$ eV. 

\subsection{Construction of the Models}

The energy barrier $E_B$ for a certain process is the difference
between the binding energies of the hopping adatom 
(to the substrate and to adjacent adatoms) at the
initial position, 
$E_{in}$,
and at the bridge site, 
$E_{top}$,  
namely $E_B = E_{top} - E_{in}$.
On the basis of the additivity feature just demonstrated, 
we will now
express these binding energies as the sum of the
occupation states of the relevant
sites.
The first approximation for the energies gives a model (model I)
with only two parameters, that reproduces the main features
of the EAM barriers.
In order to establish the model, there are two things to
note about the parameters obtained in Sec. \ref{linearity}.
First, the values of NN binding energies at the lattice site
and bridge site are very close. This reflects the fact 
that the NN distance corresponds approximately to the minimum
potential of the two-body interaction.
Second, both these energies are much larger than the NNN binding
energy.
These two features are quite general and common to all
the metals we discuss here.
For the simplest model we will neglect the effect of the NNN atoms,
and assume a single NN binding energy, 
$\Delta E_{NN}$,
for both lattice and bridge sites.
The resulting expression for the binding energy at the initial 
(fourfold hollow) site is    
\begin{equation}
E_{in}^n = E_{in}^0 - \Delta E_{NN} \cdot (S_1 + S_3 + S_5).
\end{equation}
The energy of an isolated
atom is $E_{in}^0$.
The energy of the hopping atom when it is 
on the bridge site is given by:  
\begin{equation}
E_{top}^n = E_{top}^0 - \Delta E_{NN} \cdot (S_1 + S_2 + S_5 + S_6)
\end{equation}
where $E_{top}^0$ is the energy of an isolated 
atom on top of a bridge site.
Thus, for a given configuration the barrier, 
$E_B^n = E_{top}^n - E_{in}^n$,
for an atom to hop into an adjacent vacant site is given
in model I by: 
\begin{equation}
E_{B}^n = E_{B}^0  
      +  \Delta E_{NN} \cdot (S_3 - S_2 - S_6)
\label{barrier4}
\end{equation}
where
$E_B^0 = E_{top}^0 - E_{in}^0$
and $n$ is given by Eq. 
(\ref{barindex}).
In this model only three sites affect the energy barrier,
which can take only four different values, as the
expression in the parenthesis can be either 1, 0, -1 or -2.
Each of these four barrier values 
corresponds to one of the four groups in 
Fig. \ref{histogram}.
The parameters of this model, as well as
those of the models discussed below, 
are adjusted to best fit the EAM data.
More specificly, we  found the parameters that  best describe
the 128 EAM barriers by minimizing the sum of squares:
\begin{equation}
R=\sum_{n=0}^{127} [E_B^n(EAM) - E_B^n(Model)]^2.
\label{optimize}
\end{equation}
The values obtained for these parameters for the five
metals are shown in 
Table \ref{add_table}.
Despite its simplicity, Model I can be used to describe and
analyze the main diffusion processes: single adatom hopping,
attachment, detachment and edge diffusion. 
The barriers obtained from this model can be incorporated
in simulations to reproduce (at least qualitatively) experimental
features such us cluster mobility, island morphology
and island density \cite{Biham1998}. 

The model presented above describes only the gross
features of the diffusion process.
In order to get more quantitative results, and to
better understand the importance of the different processes, it
is necessary to further refine the model.
We will now introduce model II, in which the effect of 
NNN atoms in the initial configuration is included.
The expression for the energy at the initial site is
now
\begin{eqnarray}
E_{in} & = & E_{in}^0 - \Delta E_{NN}  \cdot (S_1 + S_3 + S_5) \nonumber \\
       &   & - \Delta E_{NNN} \cdot (S_0 + S_2 + S_4 + S_6)
\end{eqnarray}
where $ \Delta E_{NNN}$ is the reduction of the energy
due to a NNN bond.
The energy barriers are now given by
\begin{eqnarray}
E_{B}^n & = & E_{B}^0 + \Delta E_{NN} \cdot (S_3 - S_2 - S_6) \nonumber \\
        &   & + \Delta E_{NNN} \cdot (S_0 + S_2 + S_4 + S_6) 
\label{barrier3}
\end{eqnarray}

Model II accounts better for processes such as detachment, edge
diffusion and vacancy diffusion, which generally involve
NNN interactions.
In the distribution of the
barriers obtained from the model, the main groups
exhibit certain widths. 
Yet, they are still significantly
narrower than the groups of the EAM barriers. 
This is due to the fact that during the hopping process
adjacent atoms may relax within their potential well. 
Model II accounts for these effects only on average and
therefore gives rise to narrower groups. 
The values obtained as best fits of the model
parameters to the EAM data
for the different metals are shown in
Table \ref{add_table}.
Further refinement can be obtained by introducing 
a distinction between the NN bond energies at the initial
four fold hollow site and that at the bridge site,
as suggested in Ref.\cite{Biham1998}.
This modification, which introduces a fourth parameter into the
model, gives only slightly better agreement with the EAM results.
In the following Section we present a more effective refinement
based on nonlinear interactions.

\subsection{Adding Non-Linear Effects}

To obtain models which provide  a better fit to
the EAM barriers,
it is necessary to consider
effects that are caused by the simultaneous interactions of
the hopping atom with several of its neighbors.
Such effects may be described by expressions such us
$S_i S_j$ or
$S_i S_j S_k$, which are equal to 1 only if all the
relevant sites are occupied, and 0 otherwise.
Such expressions are clearly beyond the linear
bond counting scheme of the previous Section.
There is a large number of possible nonlinear
interaction terms.
Our analysis of the 
EAM calculations, however, indicates that two 
of them are most significant.
The first term is related to the
shape of the diffusion path. 
It corresponds to configurations in which sites
adjacent to the bridge site, are occupied on
both sides of the diffusion path
(namely, at least one of the sites 1 and 2, as well as 
at least one of the sites 5 and 6 are occupied).
It appears that in these cases
the energy barrier is
considerably higher than for
configurations where sites on only one side of
the path are occupied.
This effect is due to the ``stiffness'' of the diffusion
path induced by the attraction from two opposite directions.
Even though there are  
nine
different such configurations,
they all contribute about the same energy difference,
and hence can be bound to a single parameter $\Delta E_{opp}$
(for opposite).
The additional term that is  now added to the 
expression for the barrier is
$\Delta E_{opp}\cdot (S_{1,2}\cdot S_{5,6})$ where $S_{i,j}=1$
if at least one of the sites i and j is occupied, and 0 if both are
empty. 
In all the metals we checked, except Cu, this term
is much larger than the NNN bond, sometimes by an order of
magnitude.

The second nonlinear interaction term is smaller, 
and is comparable to the effect of NNN sites.
It is related to the energy of the hopping atom
in the initial site. The EAM calculations indicate that 
if the two nearest neighbor sites 
1 and 5, that are symmetric with respect
to the hopping direction are both occupied, then the initial
configuration is more tightly bound. This means that if
sites 1 and 5 are both
occupied, the energy barrier is expected to be higher.   
Consequently, the corresponding term would be:
$\Delta E_{symm}\cdot (S_1 \cdot S_5) $.
The fitted value obtained for $\Delta E_{symm}$
is very close to that obtained for the NNN binding energy
$E_{NNN}$, for all five metals. 
We thus included both contributions in the same
term, although of different physical origin, to avoid the 
need for a fifth independent parameter.

The resulting model (model III)
for the hopping energy barriers is
\begin{eqnarray}
E_{B}^n & = & E_{B}^0 + \Delta E_{NN} \cdot (S_3 - S_2 - S_6) \nonumber \\
	&   & + \Delta E_{opp} (S_{1,2} \cdot S_{5,6}) \nonumber\\
	&   & + \Delta E_{NNN} \cdot (S_0 + S_4 + S_1 S_5)
\label{longform}
\end{eqnarray}

The values obtained from the best fit for the four parameters
$E_{B}^0$, $\Delta E_{NN}$, $\Delta E_{opp}$ and $\Delta E_{NNN}$ for the 
different metals are given in 
Table \ref{non_add_table}.
There are two remarks to be made about Eq. (\ref{longform}).
First, terms such as $S_1 S_5$ are 
{\it not} 
in contradiction to
the assumption that only nearest and next-nearest neighbor
interactions are significant. These terms are just a manifestation
of the simultaneous interactions of, say  
the atoms in sites $S_1$ and $ S_5$,
with the hopping atom, of which they are 
both nearest-neighbors.
Second, $S_2$ and $S_6$ are not included in the
last term of Eq. (\ref{longform}), since we found that
their dominant contribution is in the nonlinear term.

\subsection{Testing the Quality of the Fit}

The quality of the fit can be viewed in 
Fig.
\ref{models}.
The numbering of the configurations is the decimal representation of the
binary
number 
$n^{\prime} = S_3\bar{S}_2\bar{S}_6S_1S_5S_0S_4$, 
where $S_i=1(0)$ if site $i$ is occupied (unoccupied),
and $\bar{S}_i$ is the opposite of $S_i$.
There are essentially 6 groups of barriers which are marked
in the figures. These groups correspond (not necessarily
in order) to the six columns
of Table \ref{eam_table}.
As can be seen, groups II(a) and II(b) 
are in the same energy range, and together form group II
in Fig. \ref{histogram}.
Similarly
groups III(a) and III(b) coincide with group III
in Fig. \ref{histogram}.
Thus, there are actually only four groups
as mentioned in Sec. \ref{eam_section}. 
Beyond this basic division, there are some 
significant differences among the metals which 
Model III seems to handle well. The most important one
is the effect of NNN atoms on the energy barrier.
It can be seen 
from Table
\ref{non_add_table} 
that for Ag and Pd this effect is 
almost negligible, while for Cu and Ni it has much greater
importance. The effective NNN binding energy for Au is
even negative.
Although it may be
possible to construct models with
the same number of parameters that would give better
agreement with EAM results for each specific metal alone,
our approach is to find the general characteristics
of diffusion mechanisms, common to different substrates.
The agreement between the EAM barriers and model III
is slightly worse for Au than for the other 
metals. This may be due to substrate relaxation effects
which are found to be more important in this metal, and
are not accounted for in the model.

\section{Discussion and Summary}

The models presented in the previous Section
help to identify the main physical mechanisms
that determine the activation energies of self-diffusion
processes on FCC(001) metal surfaces.
Although such processes may involve interactions
with many substrate and in-plane atoms, they can be well described
as the sum of few relatively simple terms.
The first term is the activation energy for 
hopping of an isolated atom. 
The main corrections are due
to nearest-neighbor in-plane atoms at the initial site,
as well as at the bridge site.
The former increase the energy barrier
while the latter decrease it by nearly the same amount.
The next contribution is due to simultaneous presence of
atoms on both sides of the hopping atom relative to the
hopping path.
This term is important in relatively
dense environments.
Its typical value is about half that of the NN binding energy. 
A third and generally much smaller 
contribution consists of NNN bonds as well as a term
associated with the simultaneous 
presence of atoms in both sites 1 and 5 (Fig. \ref{fig:env}).

Beyond the physical understanding gained by this analysis,
the models can be used to evaluate
the activation energy of any diffusion process on the
(001) surface of the metals discussed above.
The models suggest that given a set of few activation
energies (which can be obtained from EAM,
ab-initio calculations or experiments),
it is possible to extract the complete
set of activation energy barriers.
To realize model I, for example, only two parameters
are needed. They can be obtained e.g. from the 
activation energy for single adatom hopping, and
the dissociation energy of a dimer.
To estimate a barrier using model II, a third parameter
is needed, which is the NNN binding energy. 
This may be obtained if the mobility of a 
trimer is known.
The fourth parameter which is needed for model III can be 
estimated from the activation energy for detachment 
from an atomic step.
Since model III
provides an expression for the energy barriers, 
linear in the parameters, 
any four barriers  
that give rise to four linearly independent equations,
are sufficient to determine all four parameters, and
consequently all the other barriers.

The possibility to construct a full set of activation
energy barriers from a relatively small set of parameters
is especially useful for simulations.
Without this knowledge, some processes have to be
discarded from the simulations as unimportant, or assigned 
activation energies which are not fully substantiated.
These approaches take much of the power of computer simulations,
and deny the possibility of direct quantitative 
confrontation with experimental data.
Even if a list of all relevant activation energies is
available, the model can be used to check the 
self-consistency of the data.
It can also help to interpret simulation results, which 
depend otherwise on a huge number of parameters.

The models presented here apply for diffusion on flat surfaces 
and do not describe the motion up/down steps. Such inter-terrace
moves involve a large number of possible local enviroments,
including flat steps as well as kink sites. We believe that
the approach proposed here can be extended to describe these
processes as well.

In summary, we have constructed
a family of models which describe self-diffusion on
FCC(001) metal surfaces and tested them for
Cu, Ag, Au, Ni and Pd.
For each one of these metals, 
the parameters of the models were optimized 
by comparing the energy barriers
to a full set of barriers obtained 
from semi-empirical potentials
via the embedded atom method.  
It is found that these models, 
with at most four parameters,
provide a good description 
of the hopping energy barriers
on the FCC(001) surfaces. 

We thank G. Vidali for helpful discussions.

\onecolumn


\begin{table}
\caption{
The hopping energy barriers for Cu, Ag, Au, Ni and Pd
obtained from the EAM calculations
for all possible configurations
within a $3 \times 3$ square around the hopping atom. 
The barriers
are given in eV.
Each number in the Table is the barrier $E_B^n$ for the
configuration in which the occupied sites are the union of the
occupied sites in the picture on top of the given column 
(indexed by $n_1$)
and on the left hand side of the given row
(indexed by $n_2$). 
Consequently, the index 
$n$ specifying the barrier is given by $n=n_1 + n_2$.
}
\label{eam_table}
\end{table}

\begin{table}
\caption{
The parameters 
$E_0$, 
$\Delta E_{NN}$ and
$\Delta E_{NNN}$
of model II obtained from the best fit of the 
EAM barriers for Cu, Ag, Au, Ni and Pd.
The values in parenthesis are the corresponding values
for model I. $R$ is the sum of squares defined in Eq.
(\ref{optimize}),
for the optimized parameters.
}
\label{add_table}
\end{table}

\begin{table}
\caption{
The parameters 
$E_0$, 
$\Delta E_{NN}$,
$\Delta E_{NNN}$ and
$\Delta E_{opp}$
of model III obtained from the best fit of the 
EAM barriers for Cu, Ag, Au, Ni and Pd.
$R$ is the sum of squares defined in Eq.
(\ref{optimize}).
}
\label{non_add_table}
\end{table}


\setcounter{table}{0}
\newcommand{\moveold}{
        \put(0,110){\line(1,0){30}}
        \put(0,120){\line(1,0){30}}
        \put(0,130){\line(1,0){30}}
        \put(0,140){\line(1,0){30}}
        \put(0,140){\line(0,-1){30}}
        \put(10,140){\line(0,-1){30}}
        \put(20,140){\line(0,-1){30}}
        \put(30,140){\line(0,-1){30}}
        \put(15,125){\circle{7}}
        \put(19,125){\vector(1,0){6}}
}
\newcommand{\move}{
        \put(0,160){\line(1,0){30}}
        \put(0,170){\line(1,0){30}}
        \put(0,180){\line(1,0){30}}
        \put(0,190){\line(1,0){30}}
        \put(0,190){\line(0,-1){30}}
        \put(10,190){\line(0,-1){30}}
        \put(20,190){\line(0,-1){30}}
        \put(30,190){\line(0,-1){30}}
        \put(15,175){\circle{7}}
        \put(19,175){\vector(1,0){6}}
}
\newcommand{\lzero}{
\begin{picture}(25,25)(-15,122)
        \move
\end{picture}
}
\newcommand{\lone}{
\begin{picture}(25,25)(-15,122)
        \move
        \put(5,185){\circle{7}}
\end{picture}
}
\newcommand{\ltwo}{
\begin{picture}(25,25)(-15,122)
        \move
        \put(15,185){\circle{7}}
\end{picture}
}
\newcommand{\lthree}{
\begin{picture}(25,25)(-15,122)
        \move
        \put(5,185){\circle{7}}
        \put(15,185){\circle{7}}
\end{picture}
}
\newcommand{\lsixteen}{
\begin{picture}(25,25)(-15,122)
        \move
        \put(5,165){\circle{7}}
\end{picture}
}
\newcommand{\lseventeen}{
\begin{picture}(25,25)(-15,122)
        \move
        \put(5,185){\circle{7}}
        \put(5,165){\circle{7}}
\end{picture}
}
\newcommand{\leighteen}{
\begin{picture}(25,25)(-15,122)
        \move
        \put(15,185){\circle{7}}
        \put(5,165){\circle{7}}
\end{picture}
}
\newcommand{\lnineteen}{
\begin{picture}(25,25)(-15,122)
        \move
        \put(5,185){\circle{7}}
        \put(15,185){\circle{7}}
        \put(5,165){\circle{7}}
\end{picture}
}
\newcommand{\lthirtytwo}{
\begin{picture}(25,25)(-15,122)
        \move
        \put(15,165){\circle{7}}
\end{picture}
}
\newcommand{\lthirtythree}{
\begin{picture}(25,25)(-15,122)
        \move
        \put(5,185){\circle{7}}
        \put(15,165){\circle{7}}
\end{picture}
}
\newcommand{\lthirtyfour}{
\begin{picture}(25,25)(-15,122)
        \move
        \put(15,185){\circle{7}}
        \put(15,165){\circle{7}}
\end{picture}
}
\newcommand{\lthirtyfive}{
\begin{picture}(25,25)(-15,122)
        \move
        \put(5,185){\circle{7}}
        \put(15,185){\circle{7}}
        \put(15,165){\circle{7}}
\end{picture}
}
\newcommand{\lfortyeight}{
\begin{picture}(25,25)(-15,122)
        \move
        \put(5,165){\circle{7}}
        \put(15,165){\circle{7}}
\end{picture}
}
\newcommand{\lfortynine}{
\begin{picture}(25,25)(-15,122)
        \move
        \put(5,185){\circle{7}}
        \put(5,165){\circle{7}}
        \put(15,165){\circle{7}}
\end{picture}
}
\newcommand{\lfifty}{
\begin{picture}(25,25)(-15,122)
        \move
        \put(15,185){\circle{7}}
        \put(5,165){\circle{7}}
        \put(15,165){\circle{7}}
\end{picture}
}
\newcommand{\lfiftyone}{
\begin{picture}(25,25)(-15,122)
        \move
        \put(5,185){\circle{7}}
        \put(15,185){\circle{7}}
        \put(5,165){\circle{7}}
        \put(15,165){\circle{7}}
\end{picture}
}
\newcommand{\tsixtyeight}{
\begin{picture}(25,25)(2,110)
        \moveold
        \put(25,135){\circle{7}}
        \put(25,115){\circle{7}}
\end{picture}
}
\newcommand{\tseventysix}{
\begin{picture}(25,25)(2,110)
        \moveold
        \put(25,135){\circle{7}}
        \put(25,115){\circle{7}}
        \put(5,125){\circle{7}}
\end{picture}
}
\newcommand{\tfour}{
\begin{picture}(25,25)(2,110)
        \moveold
        \put(25,135){\circle{7}}
\end{picture}
}
\newcommand{\ttwelve}{
\begin{picture}(25,25)(2,110)
        \moveold
        \put(25,135){\circle{7}}
        \put(5,125){\circle{7}}
\end{picture}
}
\newcommand{\tzero}{
\begin{picture}(25,25)(2,110)
        \moveold
\end{picture}
}
\newcommand{\teight}{
\begin{picture}(25,25)(2,110)
        \moveold
        \put(5,125){\circle{7}}
\end{picture}
}
\newcommand{\narrow}{
\begin{picture}(25,25)(2,110)
        \put(5,125){$\downarrow$}
\end{picture}
}
\begin{center}
\large{Table I}
\end{center}
\begin{center}
\begin{table}
\begin{tabular}{|l||lc|cc|cc|c|} \hline 
                 & & Group I& Group   &   II   & Group   &   III  & Group IV \\ \hline 
\hspace{0.2in} $n_1 \rightarrow$ \hspace{0.2in} &  &  68   &   76   
&   4    &   12   &   0    &   8    \\ 
   $n_2$          & & \hspace{0.5in}  & \hspace{0.5in}   
& \hspace{0.5in}  & \hspace{0.5in}  &  \hspace{0.5in}  & \hspace{0.5in}\\ 
 \narrow          & & \tsixtyeight       & \tseventysix       
& \tfour       & \ttwelve       & \tzero       & \teight    \\ \hline \hline
            &   &  &        &        &        &        &        \\ 
            &   &  &        &        &        &        &        \\ 
       &Cu & 0.01  &  0.25  &  0.18  &  0.48  &  0.48  &  0.81  \\ 
       &Ag & 0.16  &  0.37  &  0.23  &  0.47  &  0.48  &  0.72  \\ 
  0    &Au & 0.37  &  0.64  &  0.45  &  0.72  &  0.70  &  1.02  \\ 
       &Ni & 0.06  &  0.38  &  0.25  &  0.62  &  0.63  &  1.02  \\ 
       &Pd & 0.15  &  0.48  &  0.34  &  0.70  &  0.71  &  1.08  \\ 
  \lzero   &       &        &        &        &        &        &        \\ 
            &   &  &        &        &        &        &        \\ 
          &Cu &   0.02  &  0.28  &  0.25  &  0.53  &  0.46  &  0.85  \\ 
          &Ag &   0.16  &  0.37  &  0.24  &  0.46  &  0.48  &  0.72  \\ 
  1       &Au &   0.31  &  0.54  &  0.39  &  0.61  &  0.64  &  0.85  \\ 
          &Ni &   0.09  &  0.40  &  0.30  &  0.66  &  0.69  &  1.05  \\ 
          &Pd &   0.15  &  0.47  &  0.33  &  0.69  &  0.70  &  1.07  \\ 
  \lone   &   &         &        &        &        &        &        \\ 
            &   &  &        &        &        &        &        \\ 
       &Cu & 0.02  &  0.21  &  0.18  &  0.44 (0.34)  &  0.46  &  0.74 (0.60) \\ 
       &Ag & 0.15  &  0.35  &  0.22  &  0.45         &  0.48  &  0.72  \\ 
  2    &Au & 0.30  &  0.56  &  0.38  &  0.62         &  0.79  &  1.00  \\ 
       &Ni & 0.09  &  0.35 (0.10)  &  0.26  &  0.57 (0.48)  &  0.61  &  0.95 (0.79) \\ 
       &Pd & 0.16  &  0.43 (0.38 ) &  0.33  &  0.68  &  0.74  &  1.10 (0.91) \\ 
  \ltwo    &       &        &        &        &        &        &        \\ 
            &   &  &        &        &        &        &        \\ 
      &Cu &  0.05  &  0.25        &  0.24  &  0.48  &  0.54  &  0.78 (0.65)  \\ 
      &Ag &  0.16  &  0.35        &  0.23  &  0.45  &  0.50  &  0.72   \\ 
  3   &Au &  0.27  &  0.52        &  0.36  &  0.61  &  0.77  &  1.00   \\ 
      &Ni &  0.13  &  0.35 (0.33) &  0.31  &  0.63  &  0.68  &  0.99   \\ 
      &Pd &  0.17  &  0.49        &  0.35  &  0.70  &  0.77  &  1.13   \\ 
  \lthree &   &        &        &        &        &        &        \\ 
             &   &  &        &        &        &        &        \\ 
     &Cu &  {\it 0.02 }  & {\it 0.28 } &  0.21  &  0.50  & {\it 0.46} & {\it 0.85}  \\ 
      &Ag &  {\it 0.16 }  & {\it 0.37 } &  0.24  &  0.47  & {\it 0.48} & {\it 0.72}  \\ 
  16  &Au &  {\it 0.31 }  & {\it 0.54 } &  0.38  &  0.67  & {\it 0.64} & {\it 0.85}  \\ 
      &Ni &  {\it 0.09 }  & {\it 0.40 } &  0.30  &  0.65  & {\it 0.69} & {\it 1.05}  \\ 
      &Pd &  {\it 0.15 }  & {\it 0.47 } &  0.33  &  0.69  & {\it 0.70} & {\it 1.07}  \\ 
  \lsixteen &   &        &        &        &        &        &        \\ 
      &Cu &  0.04  &  0.30  &  0.28  &  0.54  &  0.66  &  0.89  \\ 
      &Ag &  0.17  &  0.36  &  0.24  &  0.47  &  0.49  &  0.72  \\ 
  17  &Au &  0.26  &  0.57  &  0.34  &  0.60  &  0.59  &  0.83  \\ 
      &Ni &  0.13  &  0.43  &  0.35  &  0.68  &  0.75  &  1.09  \\ 
      &Pd &  0.15  &  0.48  &  0.34  &  0.70  &  0.71  &  1.07  \\ 
  \lseventeen &   &        &        &        &        &        &        \\ 
            &   &  &        &        &        &        &        \\ 
      &Cu &  0.05  &  0.27  &  0.23  &  0.49  &  0.52  &  0.80  \\ 
      &Ag &  0.16  &  0.35  &  0.22  &  0.44  &  0.49  &  0.72  \\ 
  18  &Au &  0.26  &  0.57  &  0.34  &  0.55  &  0.75  &  0.98  \\ 
      &Ni &  0.12  &  0.40  &  0.30  &  0.63  &  0.66  &  1.01  \\ 
      &Pd &  0.17  &  0.48  &  0.34  &  0.68  &  0.75  &  1.11  \\  
  \leighteen &   &        &        &        &        &        &        \\ 
            &   &  &        &        &        &        &        \\ 
      &Cu &  0.08  &  0.29  &  0.29  &  0.52  &  0.61  &  0.83  \\ 
      &Ag &  0.17  &  0.35  &  0.24  &  0.45  &  0.51  &  0.73  \\ 
  19  &Au &  0.26  &  0.54  &  0.32  &  0.64  &  0.82  &  0.99  \\ 
      &Ni &  0.16  &  0.42  &  0.36  &  0.66  &  0.74  &  1.04  \\ 
      &Pd &  0.19  &  0.50  &  0.37  &  0.72  &  0.79  &  1.15  \\ 
  \lnineteen &   &        &        &        &        &        &        \\ 
            &   &  &        &        &        &        &        \\ 
      &Cu & {\it 0.02} & {\it 0.21}        & 0.18  & 0.48        & {\it 0.46} & {\it 0.74 (0.60) } \\ 
      &Ag & {\it 0.15} & {\it 0.35}        & 0.31  & 0.54        & {\it 0.48} & {\it 0.72        } \\ 
  32  &Au & {\it 0.30} & {\it 0.56}        & 0.61  & 0.86        & {\it 0.79} & {\it 1.00        } \\ 
      &Ni & {\it 0.09} & {\it 0.35 (0.10)} & 0.32  & 0.65 (0.52) & {\it 0.61} & {\it 0.95 (0.79) } \\ 
      &Pd & {\it 0.16} & {\it 0.43 (0.38)} & 0.46  & 0.77 (0.66) & {\it 0.74} & {\it 1.10        } \\ 
  \lthirtytwo &   &        &        &        &        &          &        \\ 
            &   &  &        &        &        &        &        \\ 
      &Cu & {\it 0.05 }  & {\it 0.27 }   &  0.28  &  0.54        & {\it 0.52 } & {\it 0.80 }  \\ 
      &Ag & {\it 0.16 }  & {\it 0.35 }   &  0.32  &  0.54        & {\it 0.49 } & {\it 0.72 }  \\ 
  33  &Au & {\it 0.26 }  & {\it 0.57 }   &  0.58  &  0.81        & {\it 0.74 } & {\it 0.98 }  \\ 
      &Ni & {\it 0.12 }  & {\it 0.40 }   &  0.38  &  0.69 (0.64) & {\it 0.66 } & {\it 1.01 }  \\ 
      &Pd & {\it 0.17 }  & {\it 0.48 }   &  0.47  &  0.83        & {\it 0.75 } & {\it 1.11 }  \\ 
  \lthirtythree &  &        &        &        &        &        &        \\ 
            &   &  &        &        &        &        &        \\ 
      &Cu &  0.05  &  0.24          &  0.28  &  0.50 (0.16)  &  0.55  &  0.78 (0.37)  \\ 
      &Ag &  0.17  &  0.36 (0.26)   &  0.38  &  0.58 (0.32)  &  0.64  &  0.84 (0.45)  \\ 
  34  &Au &  0.32  &  0.59          &  0.60  &  0.84         &  1.02  &  1.20         \\ 
      &Ni &  0.12  &  0.39  (0.16)  &  0.42  &  0.72 (0.29)  &  0.75  &  1.07 (0.54)  \\ 
      &Pd &  0.20  &  0.48  (0.24)  &  0.53  &  0.83 (0.40)  &  0.94  &  1.24 (0.66)  \\ 
  \lthirtyfour &   &        &        &        &        &        &        \\ 
            &   &  &        &        &        &        &        \\ 
      &Cu &  0.08  &  0.26 (0.08)  &  0.33  &  0.49         &  0.62  &  0.81 (0.47)  \\ 
      &Ag &  0.18  &  0.36 (0.26)  &  0.39  &  0.59 (0.35)  &  0.65  &  0.84 (0.56)  \\ 
  35  &Au &  0.29  &  0.54 (0.28)  &  0.57  &  0.81 (0.54)  &  0.95  &  1.13 (0.85)  \\ 
      &Ni &  0.15  &  0.41 (0.21)  &  0.47  &  0.74 (0.37)  &  0.82  &  1.09 (0.67)  \\ 
      &Pd &  0.23  &  0.50 (0.27)  &  0.56  &  0.86 (0.48)  &  0.97  &  1.28 (0.82)  \\ 
  \lthirtyfive &   &        &        &        &        &        &        \\ 
            &   &  &        &        &        &        &        \\ 
      &Cu & {\it 0.05} & {\it 0.25}        &  0.28  &  0.51        & {\it 0.54} & {\it 0.78 (0.65)} \\ 
      &Ag & {\it 0.16} & {\it 0.35}        &  0.33  &  0.54        & {\it 0.50} & {\it 0.72 } \\ 
  48  &Au & {\it 0.27} & {\it 0.52}        &  0.53  &  0.81        & {\it 0.77} & {\it 1.00 } \\ 
      &Ni & {\it 0.13} & {\it 0.35 (0.33)} &  0.38  &  0.68 (0.57) & {\it 0.68} & {\it 0.99 } \\ 
      &Pd & {\it 0.17} & {\it 0.49}        &  0.48  &  0.83        & {\it 0.77} & {\it 1.13 } \\ 
  \lfortyeight &   &        &        &        &        &        &        \\ 
            &   &  &        &        &        &        &        \\ 
      &Cu &  {\it 0.08} &  {\it 0.29} &  0.33  &  0.56  &   {\it 0.61} & {\it 0.83}  \\ 
      &Ag &  {\it 0.17} &  {\it 0.35} &  0.34  &  0.54  &   {\it 0.51} & {\it 0.73}  \\ 
  49  &Au &  {\it 0.26} &  {\it 0.54} &  0.53  &  0.88  &   {\it 0.82} & {\it 0.99}  \\ 
      &Ni &  {\it 0.16} &  {\it 0.42} &  0.44  &  0.73  &   {\it 0.74} & {\it 1.04}  \\ 
      &Pd &  {\it 0.19} &  {\it 0.50} &  0.51  &  0.85  &   {\it 0.79} & {\it 1.15}  \\ 
  \lfortynine &   &        &        &        &        &        &        \\ 
      &Cu & {\it 0.08} & {\it 0.26 (0.08)} & 0.34  & 0.51 (0.22) & {\it 0.62} & {\it 0.81 (0.47)} \\ 
      &Ag & {\it 0.18} & {\it 0.36 (0.26)} & 0.38  & 0.58 (0.34) & {\it 0.65} & {\it 0.84 (0.56)} \\ 
  50  &Au & {\it 0.29} & {\it 0.54 (0.28)} & 0.52  & 0.83 (0.57) & {\it 0.95} & {\it 1.13 (0.85)} \\ 
      &Ni & {\it 0.15} & {\it 0.41 (0.21)} & 0.47  & 0.73 (0.36) & {\it 0.82} & {\it 1.09 (0.67)} \\ 
      &Pd & {\it 0.23} & {\it 0.50 (0.27)} & 0.55  & 0.86 (0.47) & {\it 0.97} & {\it 1.28 (0.82)} \\ 
  \lfifty &   &        &        &        &        &        &        \\ 
            &   &  &        &        &        &        &        \\ 
            &   &  &        &        &        &        &        \\ 
      &Cu &  0.13  &  0.28 (0.12) &  0.40  &  0.53 (0.36)  &  0.70  &  0.90 (0.69) \\ 
      &Ag &  0.19  &  0.36 (0.28) &  0.40  &  0.58 (0.48)  &  0.67  &  0.85 (0.72) \\ 
  51  &Au &  0.30  &  0.55 (0.44) &  0.53  &  0.83 (0.69)  &  0.91  &  1.20 (0.99) \\ 
      &Ni &  0.21  &  0.43 (0.26) &  0.53  &  0.76 (0.55)  &  0.89  &  1.12 (0.86) \\ 
      &Pd &  0.26  &  0.53 (0.34) &  0.60  &  0.89 (0.66)  &  1.02  &  1.32 (1.04) \\ 
  \lfiftyone  & &        &        &        &        &        &        \\ \hline
\end{tabular}
\end{table}
\end{center}

\newpage
\begin{center}
\large{Table II}
\end{center}

\begin{table}
\vskip-\lastskip
\begin{tabular}{cr@{}c
			r@{}c
				r@{}c
					r@{}c}	
	Metal & \multicolumn{6}{c}{Model parameters[eV]} & & 
	\\ 	\cline{2-7}
	      & \multicolumn{2}{c}{$E_0$} &
		\multicolumn{2}{c}{$\Delta E_{NN}$} &
		\multicolumn{2}{c}{$\Delta E_{NNN}$} & 
		\multicolumn{2}{c}{R} 
	\\	\hline

	Cu &  & 0.487 (0.534) &   & 0.274 (0.255)  &  & 0.027 & 
		 & 0.280 (0.345)
	\\
	Ag &  & 0.509 (0.525) &   & 0.204 (0.197) &  & 0.010  &
		 & 0.450 (0.458)
	\\
	Au &  & 0.776 (0.752) &   & 0.235 (0.244) &  & -0.014 &
		 & 1.275 (1.292)
	\\
	Ni &  & 0.645 (0.697) &   &  0.326 (0.306) &  & 0.031  & 
		 &0.444 (0.526)
	\\
	Pd &  & 0.760 (0.789) &   &  0.337 (0.325) &  & 0.017  &
		 & 0.866 (0.892)
	\\
	\\
	\end{tabular}
\end{table}

\begin{center}
\large{Table III}
\end{center}

\begin{table}
\vskip-\lastskip
\begin{tabular}{cr@{}c
			r@{}c
				r@{}c
					r@{}c	
						r@{}c}		
	Metal & \multicolumn{8}{c}{Model parameters[eV]} & & 
	\\ 	\cline{2-9}
	      & \multicolumn{2}{c}{$E_0$} &
		\multicolumn{2}{c}{$\Delta E_{NN}$} &
		\multicolumn{2}{c}{$\Delta E_{NNN}$} & 
		\multicolumn{2}{c}{$\Delta E_{opp}$} & 
		\multicolumn{2}{c}{R} 
	\\	\hline

	Cu &  & 0.474 &   & 0.258  &  & 0.044 &  & 0.011  
		 & 0.159 
	\\
	Ag &  & 0.461  &   & 0.225  &  & 0.013  &  & 0.112
		 & 0.085
	\\
	Au &  & 0.686  &   & 0.294  &  & -0.017 &  & 0.199
		 & 0.346
	\\
	Ni &  & 0.610  &   &  0.322 &  & 0.046  &  & 0.067 
		 &0.141 
	\\
	Pd &  & 0.690  &   &  0.362  &  & 0.028  &  & 0.146 
		 & 0.162
	\\
	\\
	\end{tabular}
\end{table}

\newpage

\begin{figure}
\caption{Two mechanisms for surface diffusion: (a) regular or
bridge-site hopping; (b) exchange hopping. 
Dark spheres are adatoms
and light spheres are substrate atoms.}
\label{fig:terrace_diffusion}
\end{figure}	

\begin{figure}
\caption{Classification of all possible local environments 
of a hopping atom, including seven adjacent sites. 
Each site can
be either occupied or unoccupied, 
giving rise to $2^{7}=128$ local 
environments. 
Sites 1, 3 and 5 are nearest neighbors of the original 
site while sites 1, 2, 5 and 6 are adjacent to the bridge site 
that the atom has to pass.}
\label{fig:env} 
\end{figure}

\begin{figure}
\caption{The distribution of activation energies barriers
for
Cu(001), Ag(001), Au(001), Ni(001) and Pd(001), 
obtained from the
EAM calculations combined with the molecular statics procedure. 
The columns represent the number
of local configurations,
out of the 128 configurations of 
Fig.~\ref{fig:env}, 
giving rise to energy barriers in
a certain energy range.
Four groups of moves are identified in each of the five plots,
and representative moves in each group are shown.}
\label{histogram}
\end{figure}

\begin{figure}
\caption{The procedure used to evaluate
the binding energies. 
The binding energy due to two NNN bonds of 
Cu adatoms on Cu(001) is evaluated as the
difference between the total energy 
$E_1 = -8442.043$ eV
of a configuration including three
separate adatoms (a) and the total energy 
$E_2 = -8442.106$ eV
of a configuration including three adatoms
forming two NNN bonds.}
\label{linearity1}
\end{figure}	

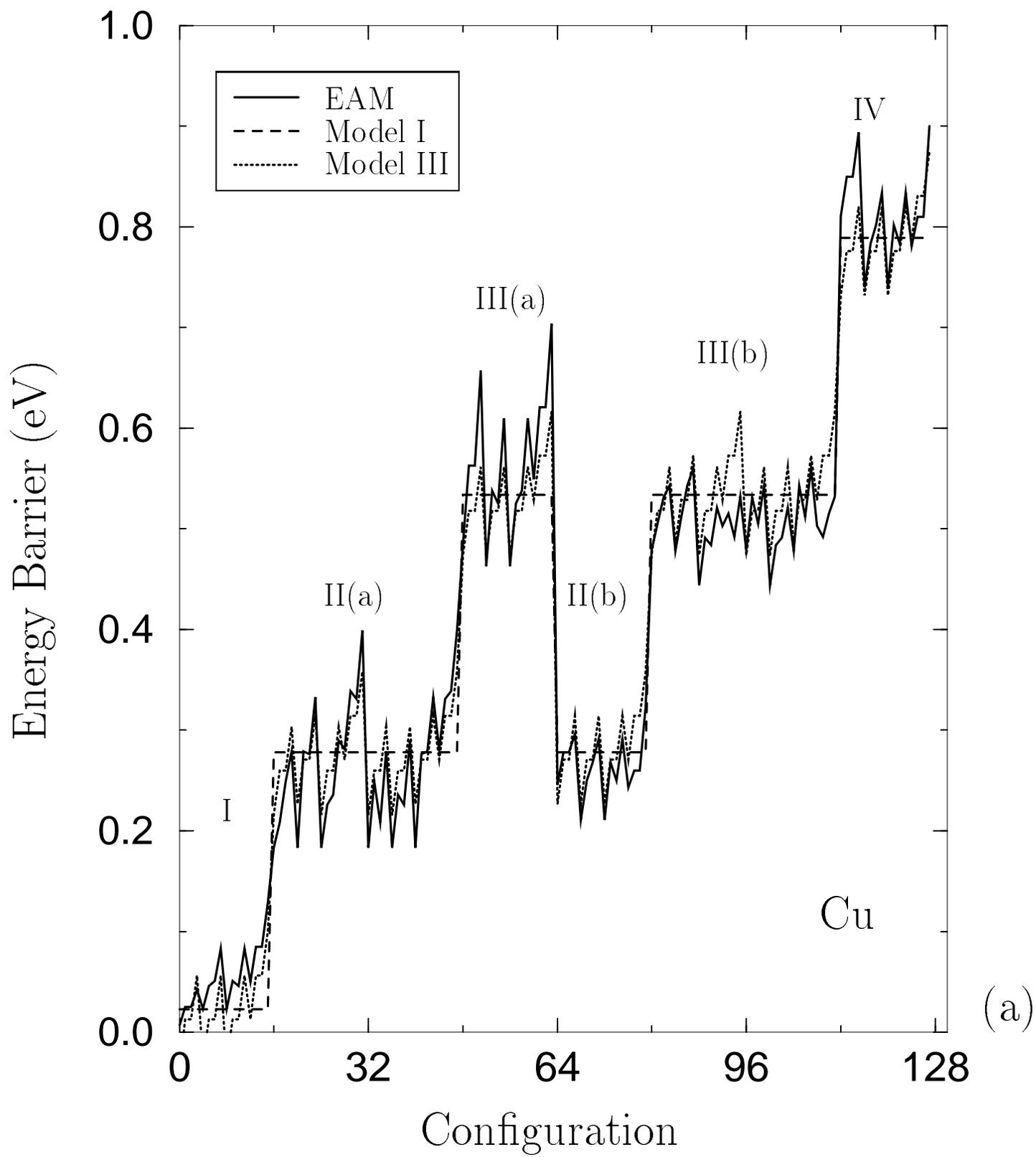
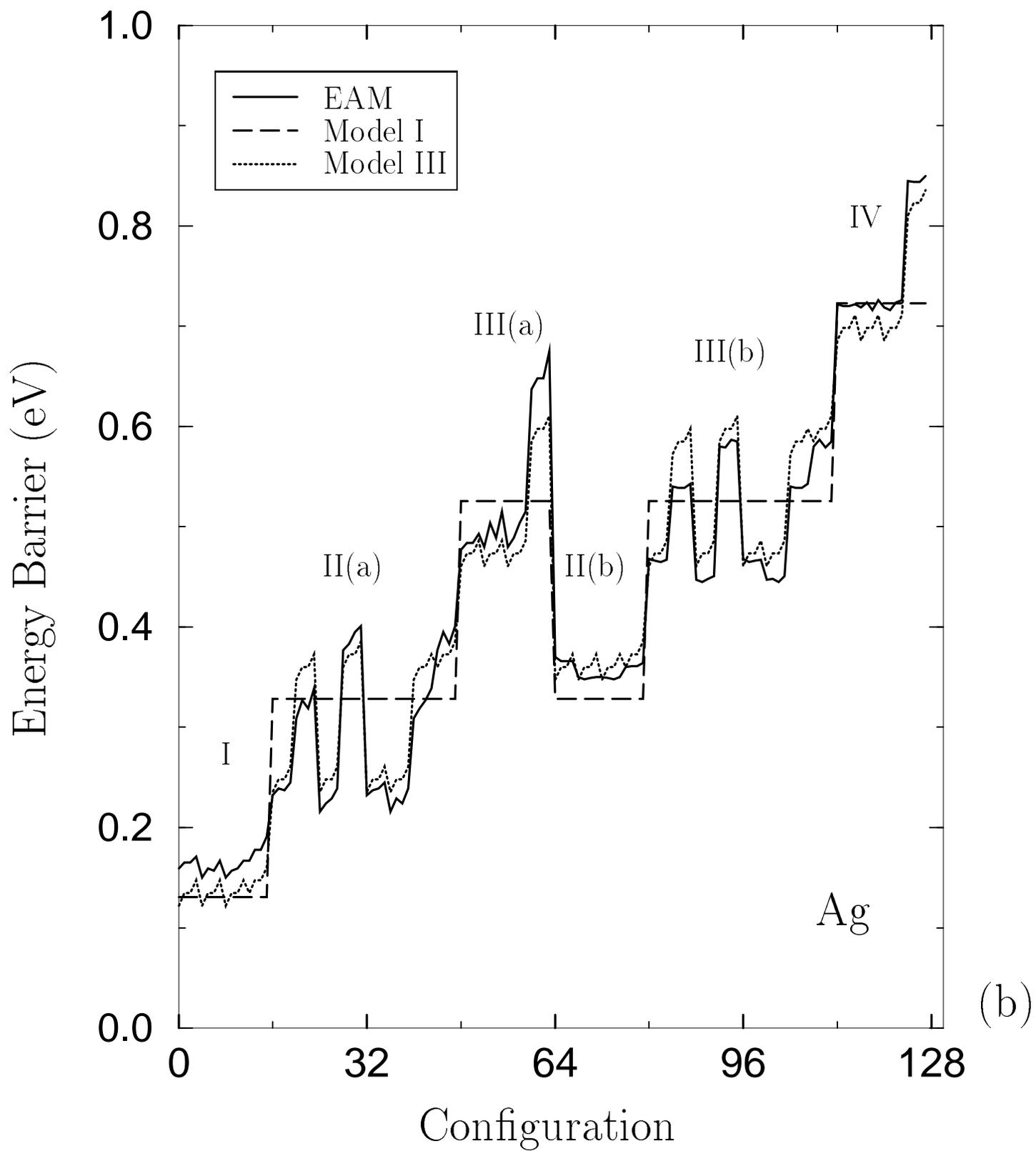
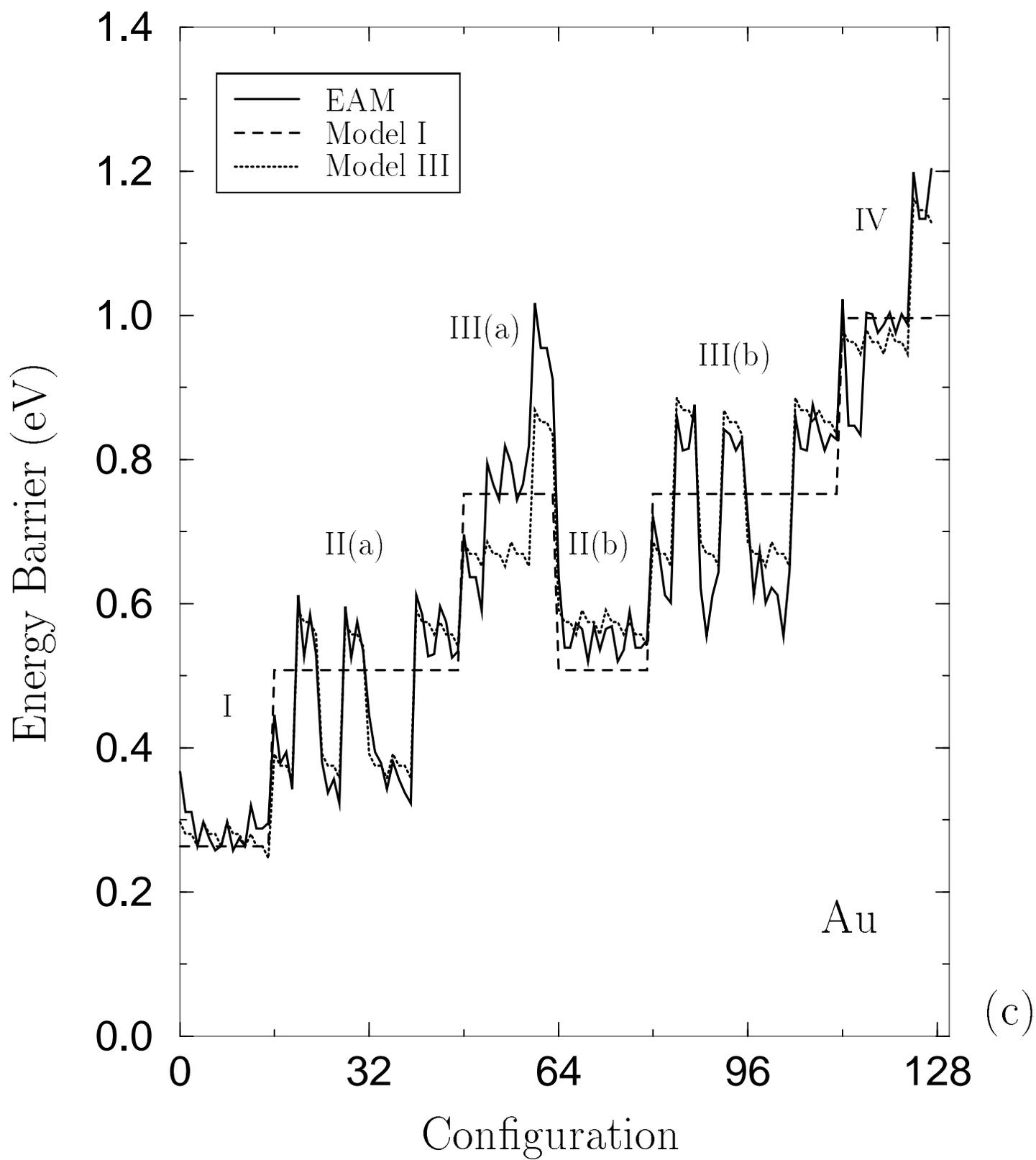
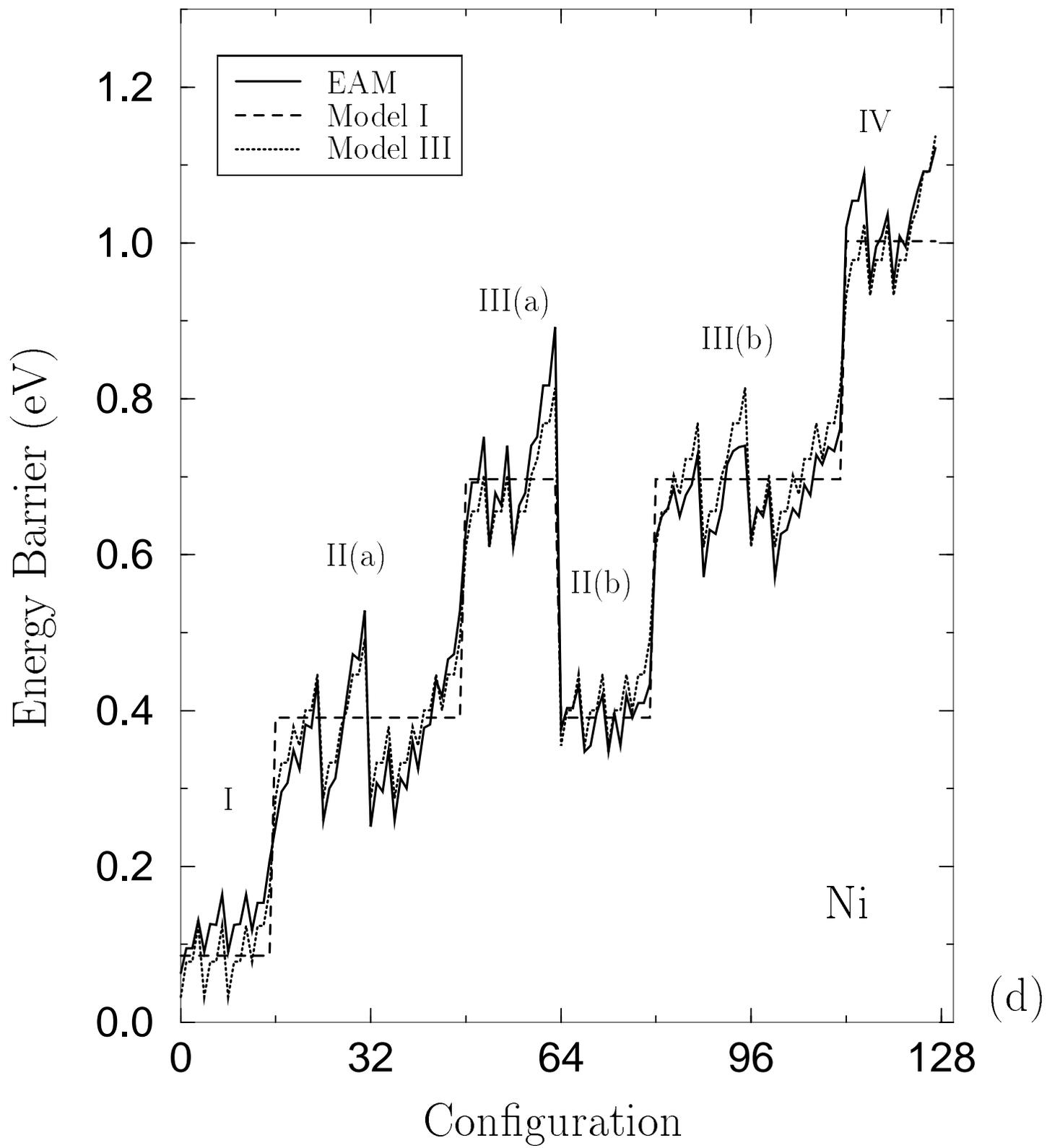
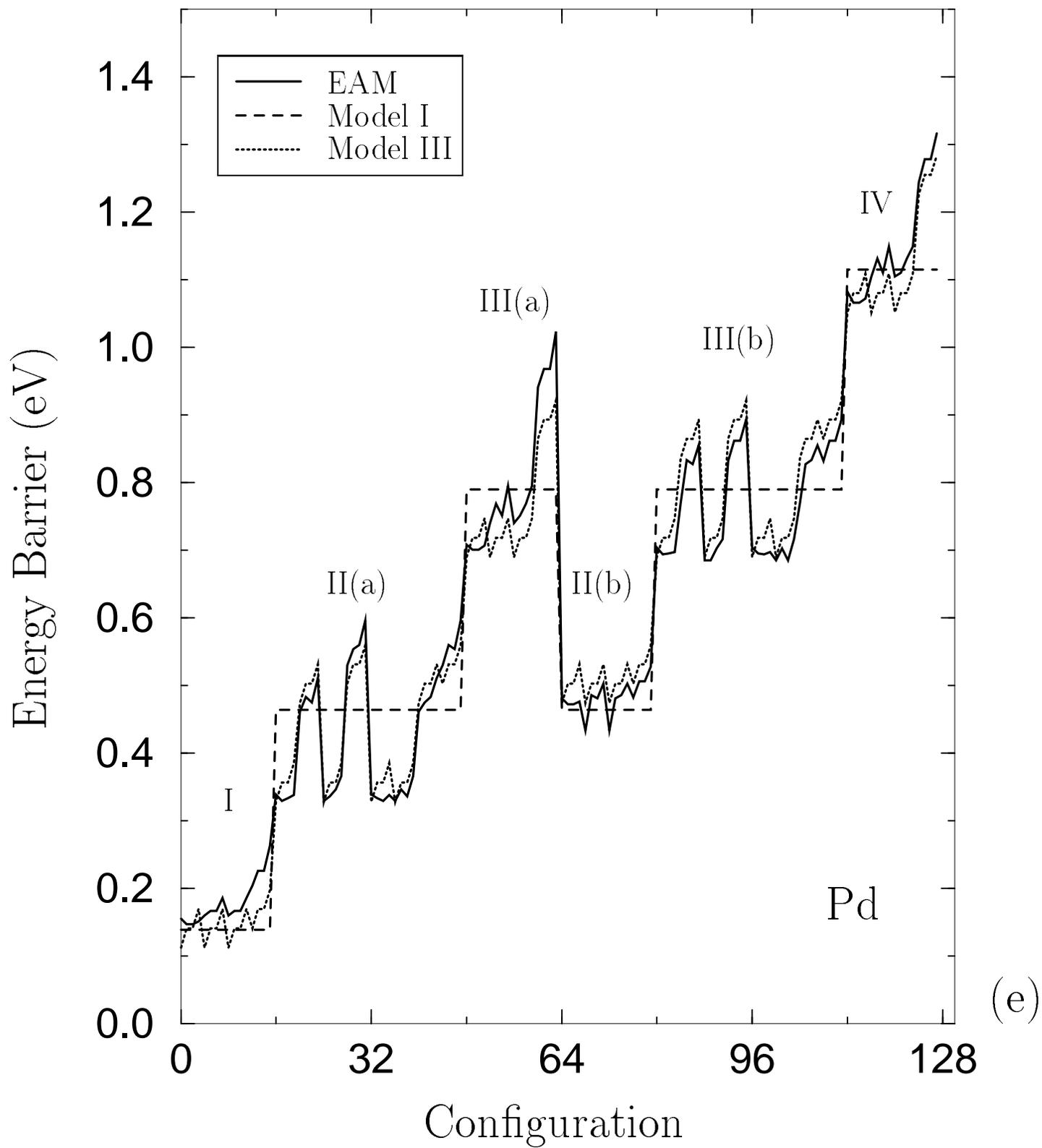
\begin{figure}
\caption{The hopping energy barriers for (a) Cu,
(b) Ag, (c) Au, (d) Ni and (e) Pd, as a function of the configuration 
number, which is given by the decimal representation of the binary
number $n^{\prime} = S_3\bar{S}_2\bar{S}_6S_1S_5S_0S_4$.  
The solid lines are the EAM energy barriers, 
the dashed lines describe the best fits obtained for model I 
and the dotted lines are the best fits for model III.
Model III is found to provide good quantitative agreement with the EAM 
results for most configurations.
There are essentially 6 groups of barriers in each plot,
which correspond 
to the six columns
of 
Table~\ref{eam_table}.
Groups II(a) and II(b) 
are in the same energy range, and together form group II
in 
Fig.~\ref{histogram},
while
groups III(a) and III(b) coincide with group III
in 
Fig.~\ref{histogram}.
}
\label{models}
\end{figure}


\newpage

\begin{thebibliography}{10}

\bibitem{Kunkel1990}
{R. Kunkel, B. Poelsema, L. K. Verheij and G. Comsa}, Phys. Rev. Lett. {\bf
  65},  733  (1990).

\bibitem{Esch1994}
{S. Esch, M. Hohage, T. Michely, and G. Comsa}, Phys. Rev. Lett. {\bf 72},  518
   (1994).

\bibitem{Ernst1992}
{H. J. Ernst, F. Fabre and J. Lapujoulade}, Phys. Rev. B {\bf 46},  1929
  (1992).

\bibitem{Ernst1992a}
{H. J. Ernst, F. Fabre and J. Lapujoulade}, Surf. Sci. {\bf 275},  {L682}
  (1992).

\bibitem{Li1993}
{W. Li, G. Vidali and O. Biham}, Phys. Rev. B {\bf 48},  8336  (1993).

\bibitem{Ernst1994}
{H.-J. Ernst, F. Fabre, R. Folkerts and J. Lapujoulade}, Phys. Rev. Lett. {\bf
  72},  112  (1994).

\bibitem{Zuo1991}
{J.-K. Zuo and J. F. Wendelken}, Phys. Rev. Lett. {\bf 66},  2227  (1991).

\bibitem{zuo1994}
{J.-K. Zuo, J. F. Wendelken, H. D\"{u}rr, and C.-L. Liu}, Phys. Rev. Lett. {\bf
  72},  3064  (1994).

\bibitem{Durr1995}
{H. D\"{u}rr, J. F. Wendelken and J.-K. Zuo}, Surf. Sci. {\bf 328},  {L527}
  (1995).

\bibitem{Swan1997}
{A. K. Swan, Z.-P. Shi, J. F. Wendelken and Z. Zhang}, Surf. Sci. {\bf 391},
  {L1205}  (1997).

\bibitem{Breeman1992}
{M. Breeman and D.O. Boerma}, Surf. Sci. {\bf {269/270}},  224  (1992).

\bibitem{Stroscio1993}
{J. A. Stroscio, D. T. Pierce and R. A. Dragoset}, Phys. Rev. Lett. {\bf 70},
  3615  (1993).

\bibitem{Hwang1991}
{R. Q. Hwang, J. Schr\"{o}der, C. G\"{u}nther and R. J. Behm}, Phys. Rev. Lett.
  {\bf 67},  3279  (1991).

\bibitem{Mo1991}
{Y. W. Mo, J. Kleiner, M. B. Webb and M. G. Lagally}, Phys. Rev. Lett. {\bf
  66},  1998  (1991).

\bibitem{Potschke1991}
{G. P\"{o}tschke, J. Schr\"{o}der, C. G\"{u}nther, R. Q. Hwang and R. J. Behm},
  Surf. Sci. {\bf {251/252}},  592  (1991).

\bibitem{Bott1992}
{M. Bott, T. Michely and G. Comsa}, Surf. Sci. {\bf 272},  161  (1992).

\bibitem{Kopatzki1993}
{E. Kopatzki, S. G\"{u}nther, W. Nichtl-Pecher and R. J. Behm}, Surf. Sci. {\bf
  284},  154  (1993).

\bibitem{Michely1993}
{T. Michely, M. Hohage, M. Bott and G. Comsa}, Phys. Rev. Lett. {\bf 70},  3943
   (1993).

\bibitem{Roder1993}
{H. R\"{o}der, E. Hahn, H. Brune, J.-P. Bucher and K. Kern}, Nature {\bf 366},
  141  (1993).

\bibitem{Girard1994}
{J. C. Girard, Y. Samson, S. Cauthier, S. Rousset and J. Klein}, Surf. Sci.
  {\bf 302},  73  (1994).

\bibitem{Gunther1994}
{S. G\"{u}nther, E. Kopatzki, M. C. Bartelt, J. W. Evans and R. J. Behm}, Phys.
  Rev. Lett. {\bf 73},  553  (1994).

\bibitem{Stroscio1994}
{J. A. Stroscio and D. T. Pierce}, Phys. Rev. B {\bf 49},  8522  (1994).

\bibitem{Vrijmoeth1994}
{J. Vrijmoeth, H.A. van der Vegt, J.A. Meyer, E. Vlieg and R.J. Behm}, Phys.
  Rev. Lett. {\bf 72},  3843  (1994).

\bibitem{Bromann1995}
{K. Bromann, H. Brune, H. R\"{o}der and K. Kern}, Phys. Rev. Lett. {\bf 75},
  677  (1995).

\bibitem{Roder1995}
{H. R\"{o}der, K. Bromann, H. Brune and K. Kern}, Phys. Rev. Lett. {\bf 74},
  3217  (1995).

\bibitem{Linderoth1996}
{T. R. Linderoth, J. J. Mortensen, K. W. Jacobsen, E. L\mbox{\ae}gsgaard, I.
  Stensgaard and F. Besenbacher}, Phys. Rev. Lett. {\bf 77},  87  (1996).

\bibitem{Wen1994}
{J.-M. Wen, S.-L. Chang, J.W. Burnett, J.W. Evans and P.A. Thiel}, Phys. Rev.
  Lett. {\bf 73},  2591  (1994).

\bibitem{Meyer1995}
{J.A. Meyer, P. Schmid and R.J. Behm}, Phys. Rev. Lett. {\bf 74},  3864
  (1995).

\bibitem{Morgenstern1995}
{K. Morgenstern, G. Rosenfeld, B. Poelsema and G. Comsa}, Phys. Rev. Lett. {\bf
  74},  2058  (1995).

\bibitem{Morgenstern1996}
{K. Morgenstern, G. Rosenfeld and G. Comsa}, Phys. Rev. Lett. {\bf 76},  2113
  (1996).

\bibitem{Pai1997}
{W.W. Pai, A.K. Swan, Z. Zhang and J.F. Wendelken}, Phys. Rev. Lett. {\bf 79},
  3210  (1997).

\bibitem{Chen1990}
{C. Chen and T. T. Tsong}, Phys. Rev. Lett. {\bf 64},  3147  (1990).

\bibitem{Kellogg1990}
{G. L. Kellogg and P. J. Feibelman}, Phys. Rev. Lett. {\bf 64},  3143  (1990).

\bibitem{Kellogg1991}
{G. L. Kellogg and A. F. Voter}, Phys. Rev. Lett. {\bf 67},  622  (1991).

\bibitem{Wang1993}
{S. C. Wang and G. Ehrlich}, Phys. Rev. Lett. {\bf 70},  41  (1993).

\bibitem{Kellogg1994}
{G. L. Kellogg}, Phys. Rev. Lett. {\bf 73},  1833  (1994).

\bibitem{Bartelt1992}
{M. C. Bartelt and J. W. Evans}, Phys. Rev. B {\bf 46},  12657  (1992).

\bibitem{Bartelt1993}
{M. C. Bartelt and J. W. Evans}, Surf. Sci. {\bf 298},  421  (1993).

\bibitem{Zhang1993}
{Z. Zhang and H. Metiu}, Surf. Sci. Lett. {\bf 292},  {L781}  (1993).

\bibitem{Amar1994}
{J. G. Amar and F. Family},  in {\em {Mechanisms of Thin Film Evolution}},
  edited by {S. M. Yalisove, C. V. Thompson and D. J. Eaglesham} ({Materials
  Research Society}, {Pittsburgh, PA, USA}, 1994), p.\ 167.

\bibitem{Bales1994}
{G. S. Bales and D. C. Chrzan}, Phys. Rev. B {\bf 50},  6057  (1994).

\bibitem{Barkema1994}
{G. T. Barkema, O. Biham, M. Breeman, D. O. Boerma, and G. Vidali}, Surf. Sci.
  Lett. {\bf 306},  {L569}  (1994).

\bibitem{Breeman1994}
{M. Breeman, G. T. Barkema and D.O.Boerma}, Surf. Sci. {\bf 303},  25  (1994).

\bibitem{Jensen1994}
{P. Jensen, A.-L. Barab\'{a}si, H. Larralde, S. Havlin and H.E. Stanley}, Phys.
  Rev. B {\bf 50},  15316  (1994).

\bibitem{Ratsch1994}
{C. Ratsch, A. Zangwill, P. Smilauer and D. D. Vvedensky}, Phys. Rev. Lett.
  {\bf 72},  3194  (1994).

\bibitem{Wolf1994}
{D. E. Wolf},  in {\em {Scale Invariance, Interfaces and Non-Equilibrium
  Dynamics}}, Vol.~344 of {\em NATO Advanced Study Institute, Series B:
  Physics}, edited by {M. Droz, A. J. McKane, J. Vannimenus and D. E. Wolf}
  (Plenum, New York, 1994), .

\bibitem{Zhang1994}
{Z. Zhang, X. Chen and M. G. Lagally}, Phys. Rev. Lett. {\bf 73},  1829
  (1994).

\bibitem{Schroeder1995}
{M. Schroeder and D. E. Wolf}, Phys. Rev. Lett. {\bf 74},  2062  (1995).

\bibitem{Amar1995}
{J. G. Amar and F. Family}, Phys. Rev. Lett. {\bf 74},  2066  (1995).

\bibitem{Ratsch1995}
{C. Ratsch, P. Smilauer, A. Zangwill and D. D. Vvedensky}, Surf. Sci. Lett.
  {\bf 329},  {L599}  (1995).

\bibitem{Bales1995}
{G. S. Bales and D. C. Chrzan}, Phys. Rev. Lett. {\bf 74},  4879  (1995).

\bibitem{Amar1996}
{J. G. Amar and F. Family}, Thin Solid Films {\bf 272},  208  (1996).

\bibitem{Breeman1996}
{M. Breeman, G. T. Barkema, M. H. Langelaar and D.O.Boerma}, Thin Solid Films
  {\bf 272},  195  (1996).

\bibitem{Furman1997}
{I. Furman and O. Biham}, Phys. Rev. B {\bf 55},  7917  (1997).

\bibitem{Voter1986}
{A.F. Voter}, Phys. Rev. B {\bf 34},  6819  (1986).

\bibitem{Clarke1988}
{S. Clarke and D.D Vvedensky}, J. Appl. Phys. {\bf 63},  2272  (1988).

\bibitem{Jacobsen1995}
{J. Jacobsen, K.W. Jacobsen, P. Stoltze and J.K. Norskov}, Phys. Rev. Lett.
  {\bf 75},  2295  (1995).

\bibitem{Biehl1998}
{M. Biehl, W. Kinzel and S. Schinzer}, Europhys. Lett. {\bf 41},  443  (1998).

\bibitem{Biham1998}
{O. Biham, I. Furman, M. Karimi, G. Vidali, R. Kennett and H. Zeng}, Surf. Sci.
  {\bf 400},  29  (1998).

\bibitem{Hansen1991}
{L. Hansen, P. Stoltze, K.W. Jacobsen and J.K. Norskov}, Phys. Rev. B {\bf 44},
   6523  (1991).

\bibitem{Hansen1993}
{L. Hansen, P. Stoltze, K.W. Jacobsen and J.K. Norskov}, Surf. Sci. {\bf 289},
  68  (1993).

\bibitem{Perkins1993}
{L.S. Perkins and A.E. DePristo}, Surf. Sci. {\bf 294},  67  (1993).

\bibitem{Liu1991}
{C.L. Liu, J.M. Cohen, J.B. Adams and A.F. Voter}, Surf. Sci. {\bf 253}, 
334 (1991)

\bibitem{Boisvert1997}
{G. Boisvert and L. J. Lewis}, Phys. Rev. B {\bf 56},  7643  (1997).

\bibitem{Karimi1995}
{M. Karimi, T. Tomkowski, G. Vidali and O. Biham}, Phys. Rev. B {\bf 52},  5364
   (1995).

\bibitem{Allnatt1993}
{A. R. Allnat and A. B. Lidiard}, {\em {Atomic Transport in Solids}}
  ({Cambridge University Press}, {Cambridge}, 1993).

\bibitem{Daw1983}
{M. S. Daw and M. I. Baskes}, Phys. Rev. Lett. {\bf 50},  1285  (1983).

\bibitem{Adams1989}
{J. B. Adams, S. M. Foiles and W. G. Wolfer}, {J. Mater. Res.} {\bf 4},  102
  (1989).

\bibitem{Foiles1986}
{S. M. Foiles, M. I. Baskes and M. S. Daw}, Phys. Rev. B {\bf 33},  7983
  (1986).

\bibitem{Finnis1984}
{M. W. Finnis and J. E. Sinclair}, Philos. Mag. A {\bf 50},  45  (1984).

\end{thebibliography}
\end{document}